\documentclass[aps,prd,reprint,twocolumn,superscriptaddress,showpacs]{revtex4-1}
\usepackage{graphicx}
\usepackage{mathrsfs}
\usepackage{bm}
\usepackage{amsmath}
\usepackage{dcolumn}
\usepackage{epstopdf}
\usepackage{dsfont}
\usepackage{amssymb}
\usepackage{tabularx}
\usepackage{array}
\usepackage{float}
\usepackage{color}
\usepackage{epstopdf}
\usepackage{mathrsfs}
\usepackage[colorlinks, linkcolor=blue,anchorcolor=blue,citecolor=blue,urlcolor=blue]{hyperref}

\begin{document}
\title{Strain-induced valley polarization, topological states, and piezomagnetism in two-dimensional altermagnetic V$_2$Te$_2$O, V$_2$STeO, V$_2$SSeO, and V$_2$S$_2$O}
\author{Jin-Yang Li}
\affiliation{School of Physics, Northwest University, Xi'an 710127, China}
\affiliation{Shaanxi Key Laboratory for Theoretical Physics Frontiers, Xi'an 710127, China}

\author{An-Dong Fan}
\affiliation{School of Physics, Northwest University, Xi'an 710127, China}
\affiliation{Shaanxi Key Laboratory for Theoretical Physics Frontiers, Xi'an 710127, China}

\author{Yong-Kun Wang}
\affiliation{School of Physics, Northwest University, Xi'an 710127, China}
\affiliation{Shaanxi Key Laboratory for Theoretical Physics Frontiers, Xi'an 710127, China}

\author{Ying Zhang}
\affiliation{Department of Physics, Beijing Normal University, Beijing 100875, China}

\author{Si Li}
\email{sili@nwu.edu.cn}
\affiliation{School of Physics, Northwest University, Xi'an 710127, China}
\affiliation{Shaanxi Key Laboratory for Theoretical Physics Frontiers, Xi'an 710127, China}

\begin{abstract}
Altermagnets (AM) are a recently discovered third class of collinear magnets, and have been attracting significant interest in the field of condensed matter physics. Here, based on first-principles calculations and theoretical analysis, we propose four two-dimensional (2D) magnetic materials--monolayer V$_2$Te$_2$O, V$_2$STeO, V$_2$SSeO, and V$_2$S$_2$O--as  candidates for altermagnetic materials. We show that these materials are semiconductors with spin-splitting in their nonrelativistic band structures. Furthermore, in the band structure, there are a pair of Dirac-type valleys located at the time-reversal invariant momenta (TRIM) X and Y points. These two valleys are connected by crystal symmetry instead of time-reversal symmetry. We investigate the strain effect on the band structure and find that uniaxial strain can induce valley polarization, topological states in these monolayer materials. Moreover, piezomagnetism can be realized upon finite doping. Our result reveals interesting valley physics in monolayer V$_2$Te$_2$O, V$_2$STeO, V$_2$SSeO, and V$_2$S$_2$O, suggesting their great potential for valleytronics, spintronics, and multifunctional nanoelectronics applications.

\end{abstract}
	
\maketitle
%\section{Introduction}
Recently, altermagnetism, a type of collinear magnetism, has attracted significant attention in condensed matter physics~\cite{vsmejkal2022beyond,vsmejkal2022emerging,bai2022observation,karube2022observation,betancourt2023spontaneous}. Altermagnetic materials possess characteristics of both ferromagnetism and antiferromagnetism, exhibiting macroscopic phenomena that break time-reversal symmetry and spin-split band structures, while maintaining antiparallel magnetic ordering without a net magnetic moment. A distinctive feature of altermagnetism is that in real space, the opposite spin sublattices are not related by spatial inversion or translation but rather by other spatial symmetries (such as fourfold rotational symmetry), and they exhibit corresponding unconventional spin polarization in reciprocal space~\cite{vsmejkal2022beyond,vsmejkal2022emerging}.
 The altermagnetic materials own variety of notable and remarkable features, such as spin-splitting and spin-momentum locking in the nonrelativistic band
structure,  the planar or bulk $d$-, $g$-, or $i$-wave symmetry of the spin-dependent Fermi surfaces, spin-degenerate nodal lines and surfaces, band anisotropy of individual spin channels~\cite{vsmejkal2022beyond,vsmejkal2022emerging}. These features in altermagnetic materials can result in many novel physical properties, such as unique spin current~\cite{bai2022observation,karube2022observation,gonzalez2021efficient}, the giant tunneling magnetoresistance~\cite{shao2021spin,hellenes2022giant},  anomalous Hall effect~\cite{feng2022anomalous} and nontrivial superconductivity~\cite{zhu2023topological}. 

To date, numerous altermagnetic materials have been discovered, leading to substantial progress in the study of this emerging field. For example, González-Hernández et al. predicted that spin-split bands in the altermagnetic material RuO$_2$ can generate spin currents ~\cite{gonzalez2021efficient}. Šmejkal et al. discovered that the chirality of RuO$_2$ crystals can give rise to the crystal Hall effect~\cite{vsmejkal2020crystal}. Zhou et al. revealed large and strongly anisotropic crystal Nernst and crystal thermal Hall effects in RuO$_2$~\cite{zhou2024crystal}. Mazin et al. predicted that FeSb$_2$ with certain alloying is an altermagnetic material exhibiting large anomalous Hall conductivity and magnetooptical Kerr effect~\cite{mazin2021prediction}. Šmejkal et al. predicted spin-polarized quasiparticles with fourfold degeneracy in the altermagnetic material CrSb~\cite{vsmejkal2022beyond}. Yuan et al. predicted significant spin splitting in the altermagnetic material MnF$_2$~\cite{yuan2020giant}. Ma et al. predicted that the giant piezomagnetism and noncollinear spin current can be realized in 2D oxide insulator V$_2$Se$_2$O ~\cite{ma2021multifunctional}. Naka et al. proposed that spin currents can be generated in $\kappa$-Cl through the application of a thermal gradient or electric field~\cite{naka2019spin}. Krempaský et al. confirmed the Kramers spin-degenerate band splitting in the altermagnetic material MnTe~\cite{krempasky2024altermagnetic}.

The characteristics of spin splitting and spin-momentum locking in altermagnetic materials also drive advancements in the field of valleytronics~\cite{schaibley2016valleytronics,vitale2018valleytronics,ma2021multifunctional}. Valley, as an emerging degree of freedom, refers to the presence of multiple energy extremal points in the Brillouin zone (BZ) for low-energy carriers in a semiconductor. Analogous to charge and spin, the valley degree of freedom can be utilized for information encoding and processing, giving rise to the concept of valleytronics~\cite{rycerz2007valley,gunawan2006valley,xiao2007valley,yao2008valley,xiao2012coupled,cai2013magnetic}. The most well-known valleytronic materials are graphene and 2D transition metal dichalcogenides. 
In these materials, the two valleys, $K$ and $K'$, are connected by time reversal symmetry ($\mathcal{T}$) and distinguished by geometric properties such as Berry curvature and orbital magnetic moment, which are odd functions under $\mathcal{T}$ symmetry~\cite{xiao2010berry}. To achieve valley polarization, a key requirement for valleytronics, the $\mathcal{T}$ symmetry must be broken. This can be done by applying a magnetic field~\cite{cai2013magnetic,li2014valley,aivazian2015magnetic,srivastava2015valley,macneill2015breaking,qi2015giant,jiang2017zeeman} or through a dynamic process like optical pumping with circularly polarized light~\cite{mak2012control,zeng2012valley,cao2012valley,hsu2015optically,mak2018light}. 

However, in altermagnetic materials, the valleys can be connected by crystal symmetry instead of time-reversal symmetry~\cite{vsmejkal2022beyond,vsmejkal2022emerging}. This unique characteristic opens up avenues for controlling valley polarization, such as through the application of strain and electric fields. For instance, recent studies have demonstrated that uniaxial strain can effectively modulate valley polarization in 2D altermagnets V$_2$Se$_2$O and V$_2$SeTeO materials~\cite{ma2021multifunctional,zhu2023multipiezo}. Similarly, electric fields have been shown to provide a predictable method for achieving valley polarization in 2D altermagnetic Ca(CoN)$_2$--family monolayers~\cite{zhang2023predictable}. Despite these promising advancements, the current inventory of 2D altermagnetic valleytronic materials remains quite limited. The scarcity of these materials poses a significant challenge for the further development and practical application of valleytronics, a field that holds immense potential for next-generation electronic and spintronic devices. Therefore, it is highly desirable to discover and explore more realistic 2D altermagnetic valleytronic materials.

In this work, based on ﬁrst-principles calculations, we predict V$_2$Te$_2$O, V$_2$STeO, V$_2$SSeO, and V$_2$S$_2$O are altermagnetic valleytronic materials. We systematically investigate their magnetic, electronic, and some interesting physical properties induced by strain, including valley polarization, topological states and piezomagnetism. We show that these materials are both dynamically and thermally stable, and have altermagnetic ground state. They are semiconductor with two valleys located at the time reversal invariant momenta (TRIM) X and Y. We show that the strain have significant impact on the band structure, and can drive these materials realize valley polarization and topological semimetal states. We find that after realized valley polarization using a strain, the hole doping can induce no zero magnetization and realize a giant piezomagnetism. It is important to emphasize that our work differs significantly from previous studies on V$_2$Se$_2$O and V$_2$SeTeO in several key aspects. Firstly, the size and type of the bandgap vary. The bandgaps of V$_2$Te$_2$O, V$_2$STeO, V$_2$SSeO, and V$_2$S$_2$O range from 0.068 eV to 1.038 eV, which not only include bandgaps comparable to those in previous studies but also encompass both smaller and larger bandgaps, as well as both direct and indirect bandgaps. Secondly, while previous research on V$_2$Se$_2$O and V$_2$SeTeO primarily focused on strain-induced valley polarization, our study demonstrates that the materials we predict not only exhibit valley polarization under strain but also undergo transitions to topological semimetal states (for V$_2$Te$_2$O and V$_2$STeO) and the changes in bandgap type (from direct to indirect or vice versa) (for V$_2$SSeO and V$_2$S$_2$O).  Our work provides some concrete materials platform for the fundamental research of 2D altermagnetic materials as well as for promising valleytronics and spintronics applications.
%\section{COMPUTATION METHODS}

Our first-principles calculations were conducted within the framework of density functional theory (DFT) employing the projector augmented wave method, as implemented in the Vienna Ab initio Simulation Package (VASP)~\cite{Kresse1994,Kresse1996,PAW}. The Perdew-Burke-Ernzerhof (PBE)-type generalized gradient approximation was used for the exchange-correlation functional~\cite{PBE}. The cutoff energy was set as 600 eV. The energy and force convergence criteria were set to be $10^{-6}$ eV and 0.01 eV/\AA~respectively. The Brillouin zone (BZ) was sampled using a $\Gamma$-centered $k$ mesh of size $12\times 12\times 1$. A vacuum layer with a thickness of 20 \AA~was introduced to avoid artificial interactions between periodic images. The irreducible representations (IRs) of the bands
were calculated by the irvsp code ~\cite{gao2021irvsp}.  The correlation effects for the V-3$d$ electrons were treated by the DFT+$U$ method~\cite{anisimov1991,dudarev1998}, where the value of the Hubbard $U$ is taken as 4 eV, consistent with previous reports~\cite{zhu2023multipiezo,cheng2021intrinsic,xing2020anisotropic}. The test of
other U values for monolayer V$_2$Te$_2$O is presented in supplementary material.  The band structure results of these materials were further checked by using the Heyd-Scuseria-Ernzerhof hybrid functional method (HSE06)~\cite{heyd2003hybrid} (see the supplementary material). The ab initio molecular dynamics (AIMD) simulations were carried out using $3\times 3\times 1$ supercell at 300 K with a 1 fs time step~\cite{nose1984unified}. The phonon dispersion spectrum using density-functional perturbation theory (DFPT) with a $2\times 2\times 1$ supercell was calculated to verify the dynamic stability using the PHONOPY code~\cite{togo2015first}.

%\section{Crystal structure and stability}

\begin{figure}[htb]
	\includegraphics[width=8.6cm]{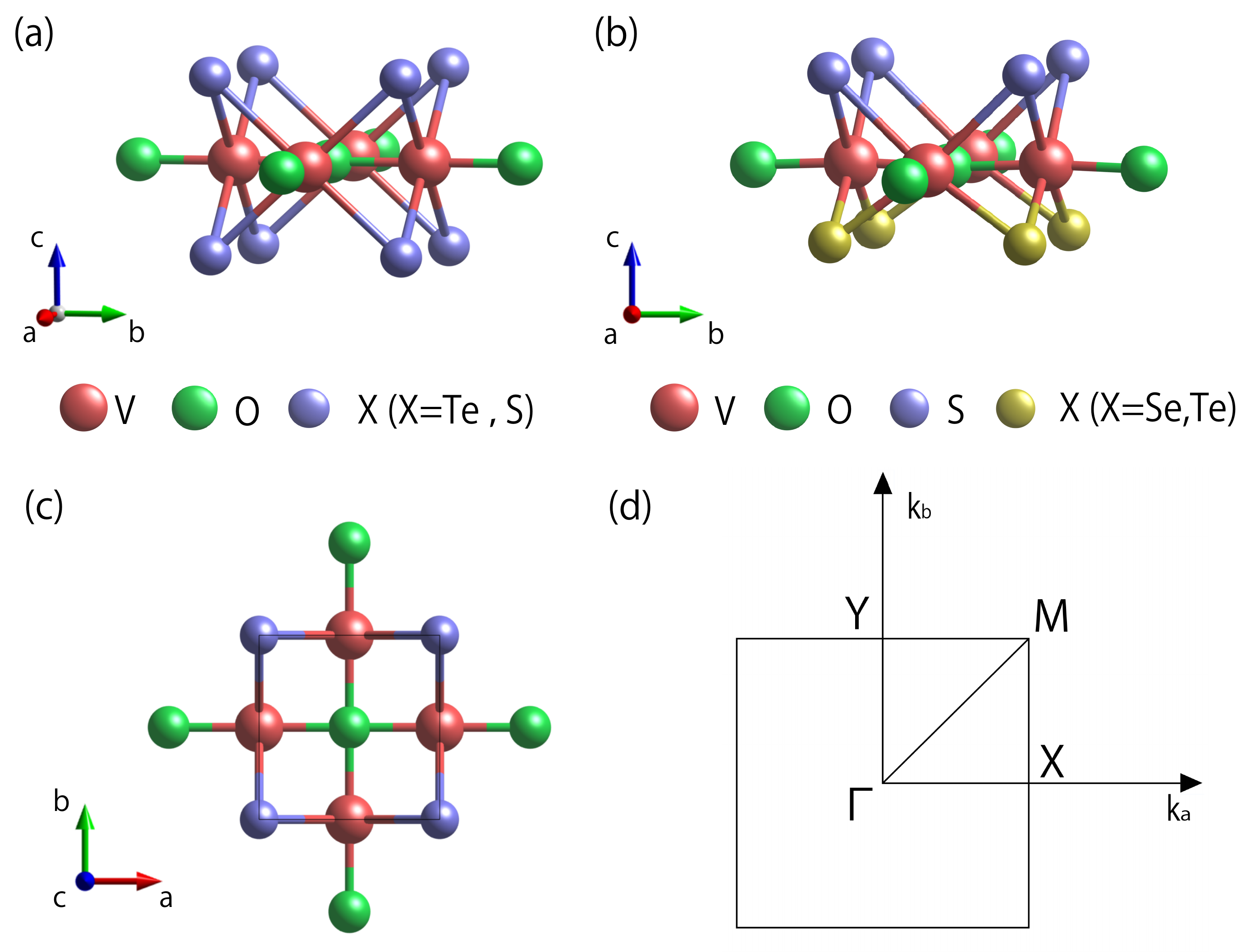}
	\caption{Side view of the crystal structure of the monolayer: (a) V$_2$Te$_2$O and V$_2$S$_2$O, (b) V$_2$STeO and V$_2$SSeO. (c) Top view of the monolayer crystal structure. (d) Brillouin zone for the monolayer structure. The high-symmetry points are labeled.}
	\label{fig1}
\end{figure}
 Single crystals of bulk V$_2$Te$_2$O have been successfully synthesized in experiments, confirming the presence of dominant interlayer van der Waals interactions~\cite{ablimit2018v2te2o}.
The monolayer V$_2$Te$_2$O and V$_2$S$_2$O are three-layer atomic structure sandwiched by two Te and S planes. They have the tetragonal lattice structure with space group $P4/mmm$ (No. 123), as shown in Fig.~\ref{fig1} (a) and (c). The monolayer V$_2$STeO and V$_2$SSeO are obtained by substituting S with Se or Te in the
bottom layer of V$_2$S$_2$O. They exhibit lower symmetry due to out-of-plane mirror symmetry breaking while retaining the tetragonal structure, and belong to space group $P4mm$ (No. 99), as shown in Fig.~\ref{fig1} (b). The optimized lattice parameters for the monolayer V$_2$Te$_2$O, V$_2$STeO, V$_2$SSeO, and V$_2$S$_2$O are $a=b=4.1$ \AA, $a=b=4.01$ \AA, $a=b=4.03$ \AA, and $a=b=4.012$ \AA, respectively.

Because the results of V$_2$SSeO, V$_2$S$_2$O and  V$_2$Te$_2$O, V$_2$STeO share very similar features, in the following discussion, we will focus on the results for V$_2$Te$_2$O, V$_2$STeO. The results for V$_2$S$_2$O and V$_2$SSeO are presented in the supplementary material. 

\begin{figure}[htb]
	\includegraphics[width=8.6cm]{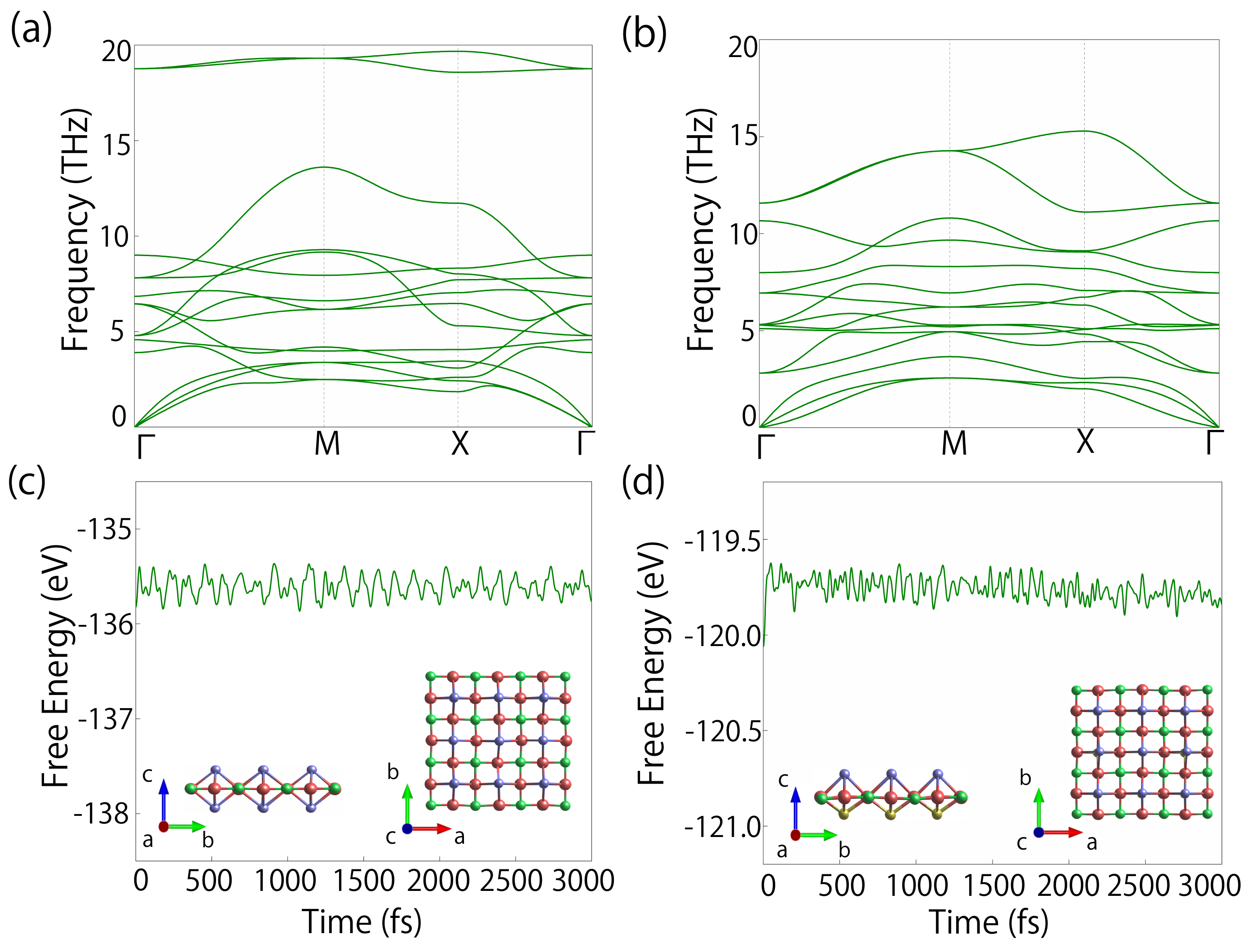}
	\caption{Calculated phonon spectrum of (a) V$_2$Te$_2$O and (b) V$_2$STeO. Molecular dynamics (MD) simulation of (c) V$_2$Te$_2$O and (d) V$_2$STeO.}
	\label{fig2}
\end{figure}

The structural stability of V$_2$Te$_2$O and V$_2$STeO are investigated by computing the phonon
spectra and by AIMD simulations. The calculated phonon spectrum for V$_2$Te$_2$O and V$_2$STeO are shown in Fig.~\ref{fig2} (a) and (b), respectively. One observes that there is no imaginary frequency in the spectrum, showing the dynamical stability of the structure.
In addition, the thermal stability of the materials are verified by the AIMD simulations. We perform the simulation for temperatures at 300 K and find that these
materials still maintain their structural integrity (see Fig.~\ref{fig2} (c) and (d)), confirming their excellent thermal stability.

%\section{RESULTS}
%\section{Magnetic configuration}
 \begin{figure}[htb]
	\includegraphics[width=8.6cm]{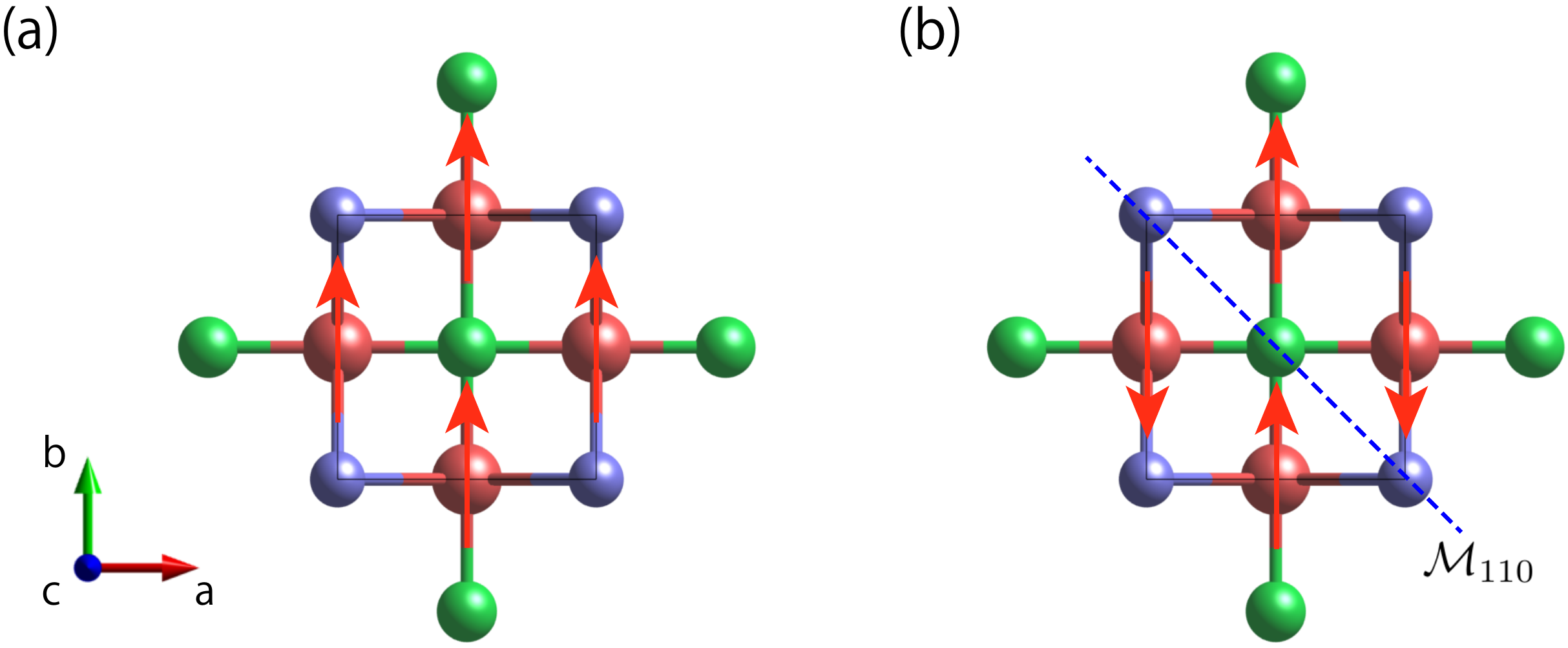}
	\caption{The magnetic conﬁgurations that we have considered: (a) is for the ferromagnetic conﬁguration, and (b) is for the AFM conﬁgurations. The red arrows represent the direction of the magnetic moments.}
	\label{fig3}
\end{figure}
Because these monolayer materials contains the $3d$ transition metal element V which usually exhibits magnetism, we ﬁrst determine the magnetic ground state for these material.  We determine their magnetic ground state by comparing the total energies of three typical magnetic configurations, including paramagnetic (NM), ferromagnetic (FM), and antiferromagnetic (AFM) configurations. The FM and AFM configurations are shown in Fig.~\ref{fig3} (a) and (b), respectively. By comparing the calculated total energies we find V$_2$Te$_2$O and V$_2$STeO  both prefer the AFM ground state.
In the AFM state, we find that magnetic moments are mainly distributed on the V site, with a large value 2 $\mu_B/V$. Interestingly, the two V atoms with opposite spins are connected by mirror symmetry ${\mathcal{M}}_{110}$. This type of AFM state is also referred to as altermagnetism (AM), enabling spin splitting in band structure~\cite{vsmejkal2022beyond,vsmejkal2022emerging}.
%\section{ELECTRONIC BAND AND VALLEY STRUCTURES}

\begin{figure}[htb]
	\includegraphics[width=8.6cm]{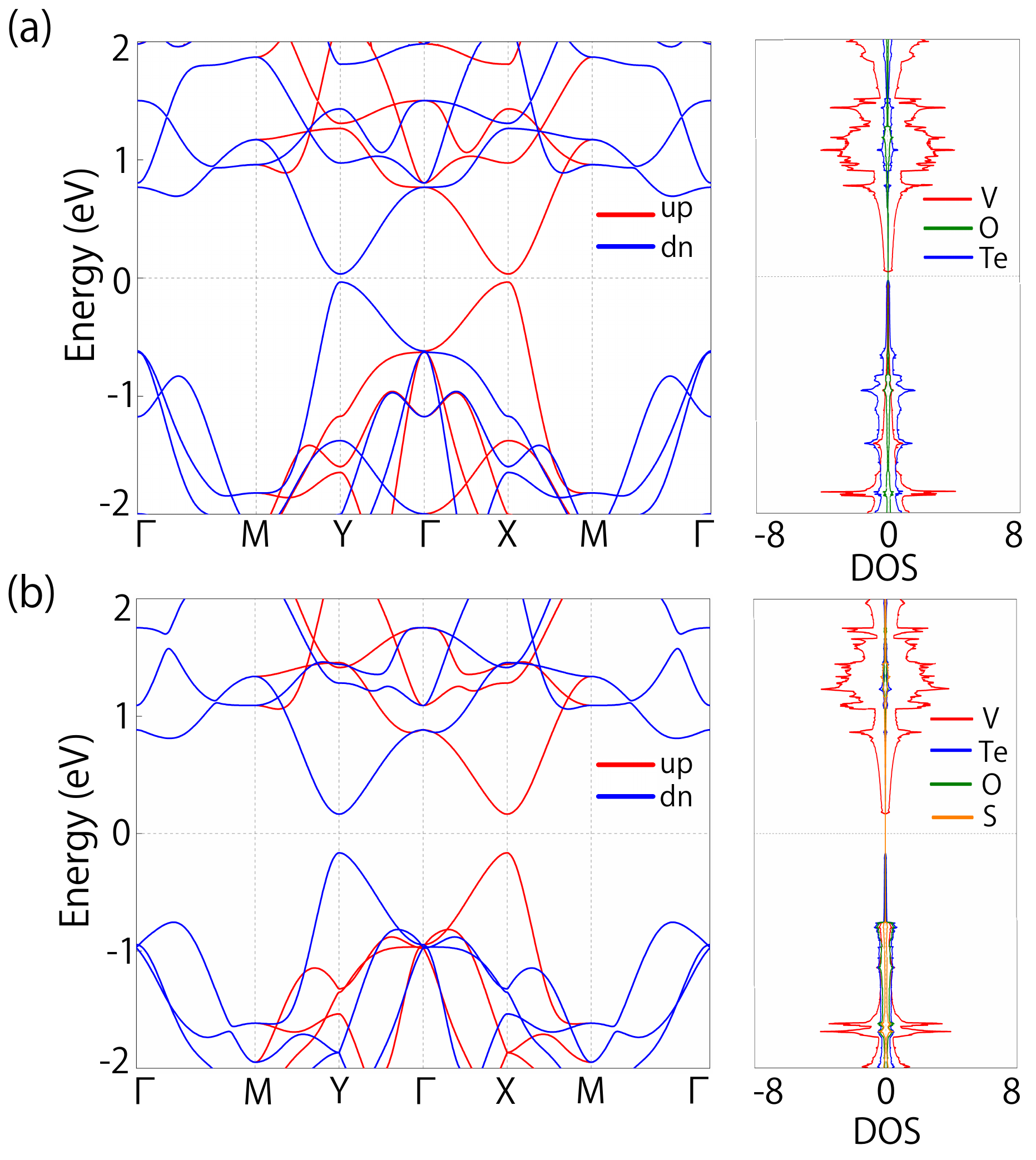}
	\caption{(a) Band structure and partial density of states (DOS) of monolayer (a) V$_2$Te$_2$O and (b) V$_2$STeO without SOC. The red (blue) color represents spin-up (spin-down) bands.}
	\label{fig4}
\end{figure}

In the following, we investigate the electronic band structure for the ground state of monolayer V$_2$Te$_2$O and V$_2$STeO (with AM). Since the spin-orbit coupling (SOC) has little effect on the band structure and valleys, we consider the band structure in the absence of SOC in the main text. The band structures of V$_2$Te$_2$O and V$_2$STeO with SOC included are shown in the supplementary material. The band structure and density of states (DOS) without SOC from our calculation for monolayer V$_2$Te$_2$O and V$_2$STeO are shown in Fig.~\ref{fig4} (a) and (b), respectively. One observes that monolayer V$_2$Te$_2$O and V$_2$STeO both are direct band gap semiconductors with the conduction band minimum (CBM) and valence band maximum (VBM) located at TRIM X and Y points. The band
gap values obtained with the PBE method are about 0.068 eV for V$_2$Te$_2$O and 0.72 eV for V$_2$STeO. The band gap of V$_2$Te$_2$O is smaller than that of V$_2$Se$_2$O (0.72 eV), while the band gap of V$_2$STeO is larger than that of V$_2$SeTeO (0.25 eV)~\cite{ma2021multifunctional,zhu2023multipiezo}.
We also used the more sophisticated hybrid functional method (HSE06)~\cite{heyd2003hybrid} to verify the band structure results. The main band features were found to be consistent between the two methods, with the band gap values from the HSE06 approach being 0.3095 eV for V$_2$Te$_2$O and 0.60 eV for V$_2$STeO. The low-energy states near the Fermi level are mainly from the V atoms. The CBM and VBM of band structure formed two valleys at X and Y points, and the two valleys are degenerate and connected by the mirror symmetry ${\mathcal{M}}_{110}$. Moreover, the band structures in these materials exhibit obvious spin-splitting and there are different spin at valley X and valley Y, formed spin-valley locking (SVL) without SOC. This arises from its altermagnetic order shown in Fig.~\ref{fig3} (b), and are allowed by the spin-group symmetry~\cite{vsmejkal2022emerging}.

%\section{Strain-induced valley polarization and topological states}
 \begin{figure*}[htb]
	\includegraphics[width=16.8cm]{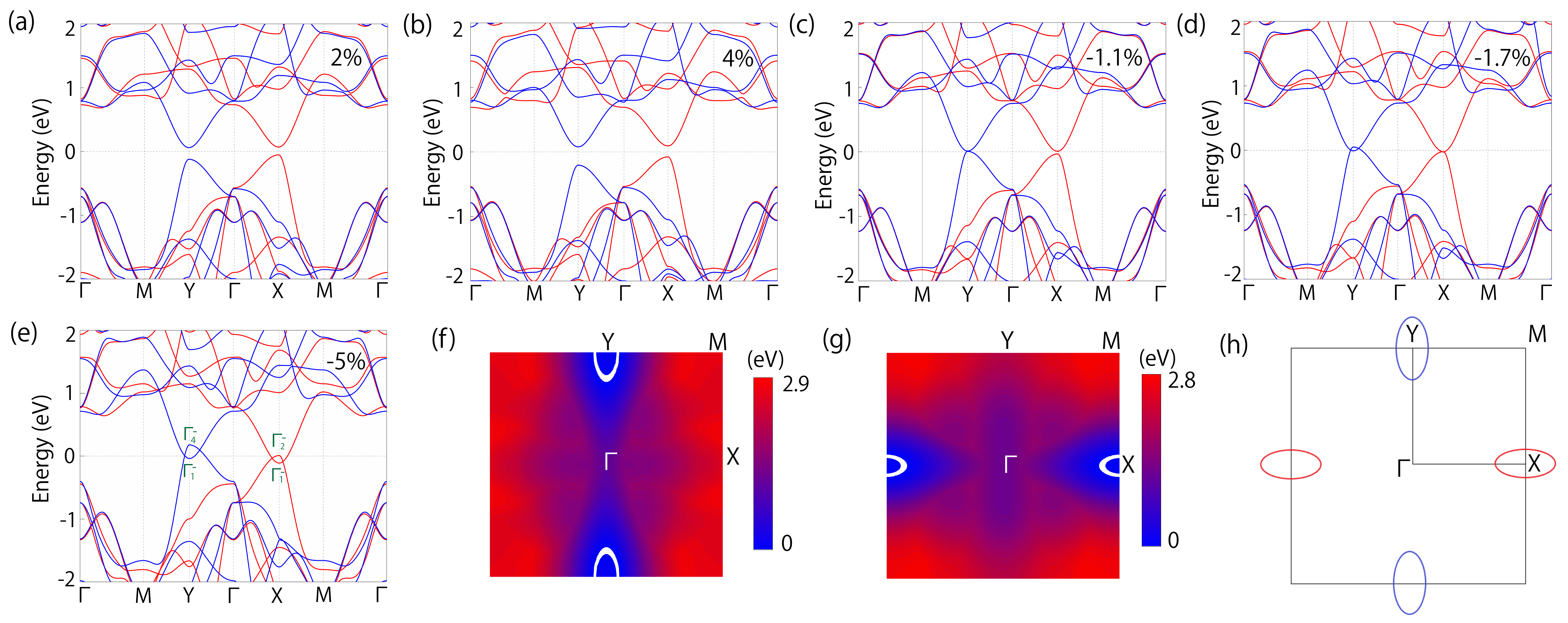}
	\caption{(a)-(e) Band structure evolution under different uniaxial strain along the $a$ direction of monolayer V$_2$Te$_2$O. Shape of the nodal loop (the white-colored
		loop) formed by the (f) spin-down bands and (g) spin-up bands under $-5\%$ strain, obtained from the DFT calculations. The color map indicates
		the local gap between the two crossing bands. (h) Schematic figure showing the nodal loops on the Brillouin zone (BZ).}
	\label{fig5}
\end{figure*}

\begin{figure}[htb]
	\includegraphics[width=8.6cm]{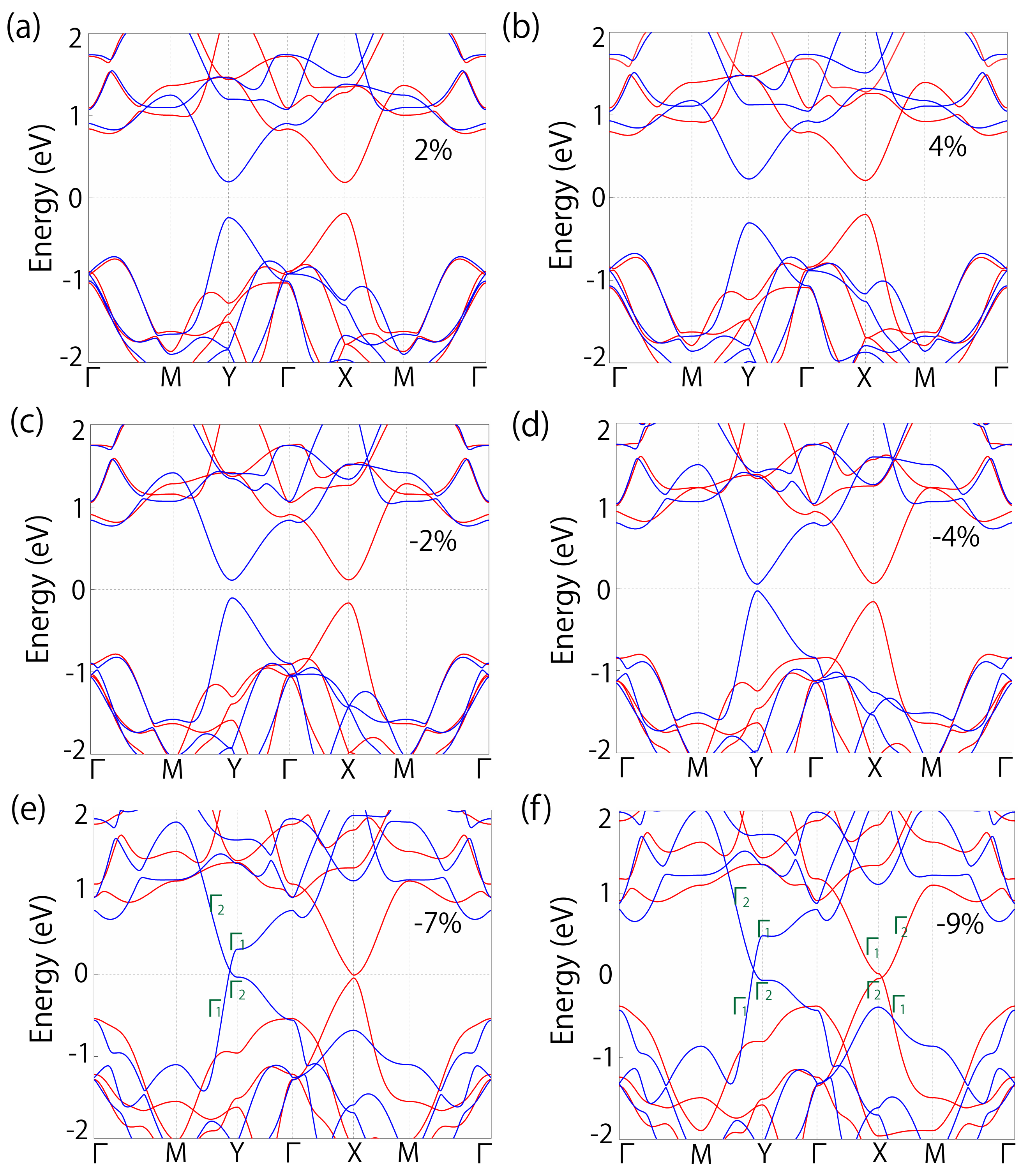}
	\caption{(a)-(f) Band structure evolution under different uniaxial strains along the $a$ direction of monolayer V$_2$STeO. The irreducible representations (IRRs) are given in (e) and (f).}
	\label{fig6}
\end{figure}

Since the two valleys at X and Y are degenerate and connected by the mirror symmetry ${\mathcal{M}}_{110}$ instead of time reversal symmetry, one would expect to break the degeneracy of the two valleys and realize the valley polarization by breaking the mirror symmetry ${\mathcal{M}}_{110}$. In fact, we can break the mirror symmetry ${\mathcal{M}}_{110}$ by applying uniaxial strain. The band structures evolution under different uniaxial strains along $a$ direction for the monolayer V$_2$Te$_2$O and V$_2$STeO materials are shown in Fig.~\ref{fig5} and Fig.~\ref{fig6}, respectively. We have also verified the magnetic ground state and stable of the monolayer under the strain (see the supplementary material). In the following, we will analyze the strain effect on the band structures of monolayer V$_2$Te$_2$O and V$_2$STeO materials in detail. 
    
\begin{figure}[htb]
	\includegraphics[width=8.6cm]{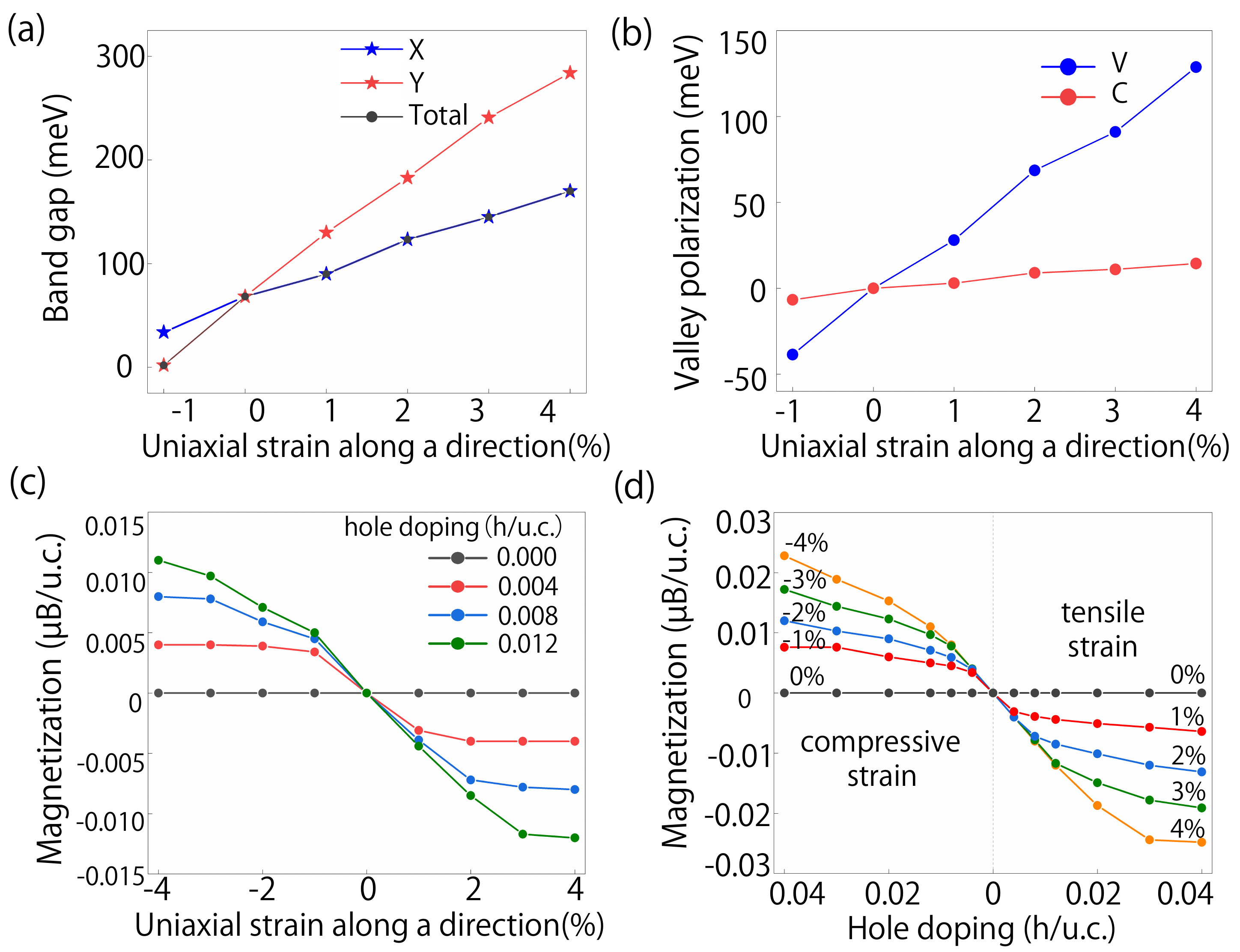}
	\caption{(a) Valley gaps at X and Y, as well as the total
		bandgap of monolayer V$_2$Te$_2$O. (b) The strain-controlled valley polarization $P=E(X)-E(Y)$ generated at the valence band (blue) and the
		conduction band (red) of monolayer V$_2$Te$_2$O. (c) and (d) The corresponding net magnetization per unit cell for different concentrations of hole doping.}
	\label{fig7}
\end{figure}

For the monolayer V$_2$Te$_2$O, we investigate the effects of uniaxial strain ranging from $-5\%$ to 4$\%$, as shown in Fig.~\ref{fig5}. Several transitions in the band structure can be observed. Under uniaxial tensile strain, the global band gap increases, and the local band gap at the Y point becomes larger than that at the X point, leading to valley polarization. This breaking of valley degeneracy arises from the breaking of mirror symmetry ${\mathcal{M}}_{110}$ due to the uniaxial tensile strain. Conversely, with uniaxial compressive strain, the global band gap decreases, and the local band gap at the Y point becomes smaller than that at the X point, also resulting in valley polarization.
The band gap variations at the X and Y points, as well as the global band gap, are shown in Fig.~\ref{fig7} (a). The valley polarization can be defined as the energy difference between two valleys $P = E(X)-E(Y)$~\cite{ma2021multifunctional}. The valley polarization of valence and conduction bands versus
uniaxial strain along the $a$ direction is shown in Fig.~\ref{fig7} (b). When the uniaxial compressive strain exceeds $-1.1\%$, the gap closes for the spin-down band at the Y point, while the band gap for the spin-up band at the X point remains, leading to the formation of a Weyl point with a specific spin channel at the Y point [see Fig.~\ref{fig5}(c)]. At strains above $-1.7\%$, two band crossing points form at the Y point, which a careful scan shows to be located on a nodal loop around the Y point. Simultaneously, the band gap at the X point closes, forming a Weyl point [see Fig.~\ref{fig5}(d)]. With further increases in uniaxial compressive strain, such as $-5\%$, the bands near the Fermi level form nodal loops at both the Y and X points, each with different spin channels [see Fig.~\ref{fig5}(e)-(h)]. Symmetry analysis reveals that the nodal loops around the Y and X points are protected by the mirror symmetry ${\mathcal{M}}_{z}$. The nodal loops lie in the 2D plane, which is invariant under ${\mathcal{M}}_{z}$. Hence, each state in this plane is also an eigenstate of ${\mathcal{M}}_{z}$, with a well-defined eigenvalue $m_z=\pm 1$. The nodal loops are protected if the two crossing bands have opposite $m_z$ in the  plane, which is indeed the case as verified by our DFT calculations [as shown in Fig.~\ref{fig5} (e)]. The IRs of the two bands near the Fermi level at X point are $\Gamma_{1}^{-}$ and $\Gamma_{2}^{-}$, and at Y point are $\Gamma_{1}^{-}$ and $\Gamma_{4}^{-}$, which correspond to the opposite $m_z$ eigenvalue.

\begin{figure}[htb]
	\includegraphics[width=8.6cm]{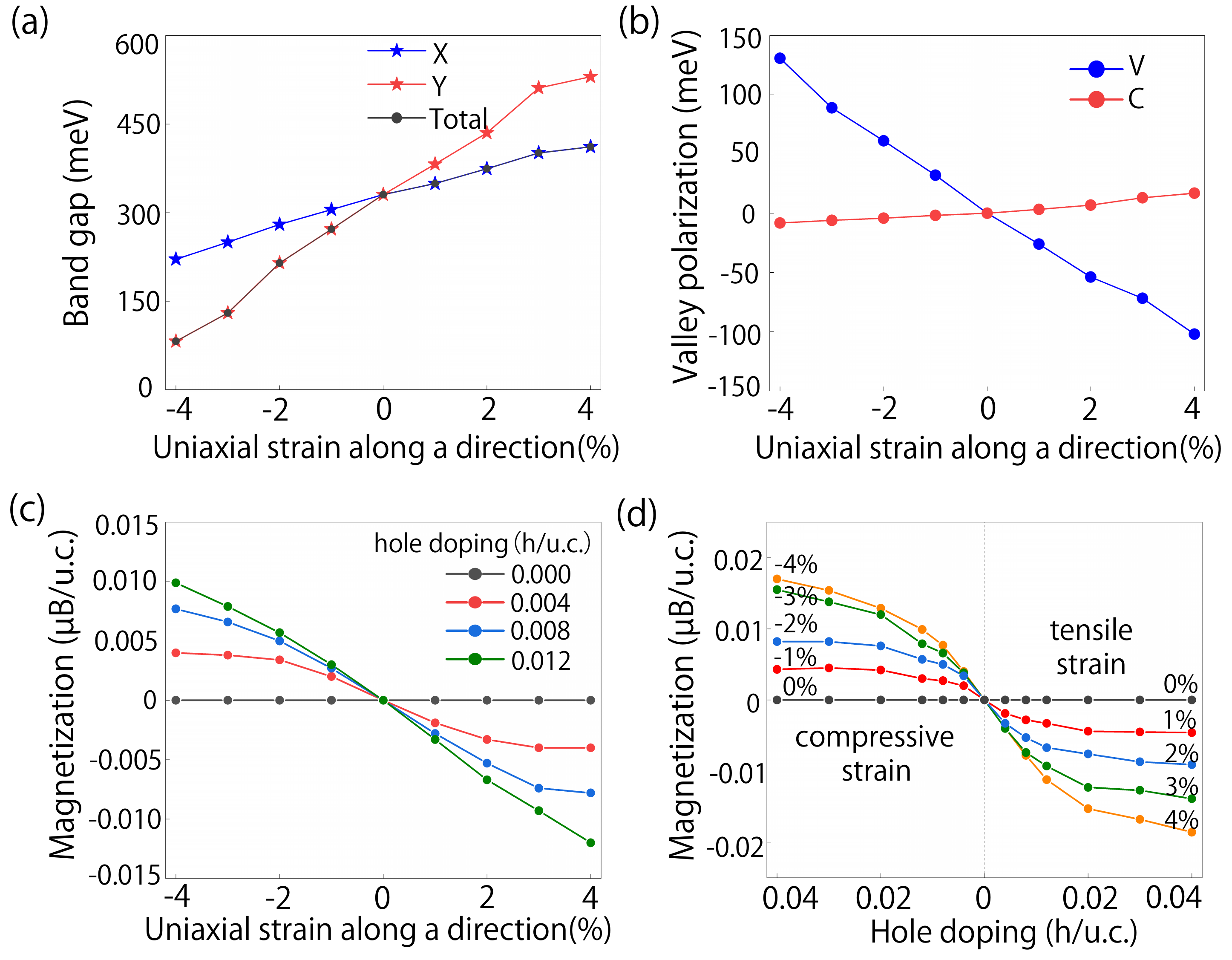}
	\caption{(a) Valley gaps at X and Y, as well as the total
		bandgap of monolayer V$_2$STeO. (b) The strain-controlled valley polarization $P=E(X)-E(Y)$ generated at the valence band (blue) and the
		conduction band (red) of monolayer V$_2$STeO. (c) and (d) The corresponding net magnetization per unit cell for different concentrations of hole doping.}
	\label{fig8}
\end{figure}

For the monolayer V$_2$STeO, we investigate the effects of uniaxial strain ranging from $-9\%$ to 4$\%$, as shown in Fig.~\ref{fig6}. Under uniaxial tensile strain, monolayer V$_2$STeO exhibits similar behavior to monolayer V$2$Te$2$O: the global band gap increases, and the local band gap at the Y point becomes larger than that at the X point, resulting in valley polarization. When uniaxial compressive strain exceeds $-4\%$, the global band gap decreases, and the local band gap at the Y point becomes smaller than that at the X point, again leading to valley polarization. The band gap variations at the X and Y points, as well as the global band gap, are shown in Fig.~\ref{fig8} (a). The valley polarization of valence and conduction bands versus
uniaxial strain along the $a$ direction is shown in Fig.~\ref{fig8} (b). However, when the uniaxial compressive strain reaches $-7\%$, the spin-down band forms a Weyl point near the Y point. With further increases in uniaxial compressive strain, such as $-9\%$, Weyl points form at both the Y and X points, each with different spin channels [see Fig.~\ref{fig6}(f)]. This is different from the Weyl points in ferromagnetic Weyl semimetals, where there is only one spin channel~\cite{you2019two,li2021correlation,tan2024bipolarized}. Symmetry analysis shows that the Weyl points near the Y and X points are protected by the mirror symmetries ${\mathcal{M}}_{y}$ and ${\mathcal{M}}_{x}$, respectively. The Weyl point near the Y point (X point) located on the path Y-M: $(k_x,\pi)$ [X-M: $(\pi,k_y)$], which is an invariant subspace of the symmetry${\mathcal{M}}_{y}$ (${\mathcal{M}}_{x}$). 
 Hence, each state on this path is also an eigenstate of ${\mathcal{M}}_{y}$ (${\mathcal{M}}_{x}$), with a well-defined eigenvalue $m_y=\pm 1$ ($m_x=\pm 1$). The Weyl point is protected if the two crossing bands have opposite $m_y$ ($m_x$) on the Y-M (X-M) path, a condition confirmed by our DFT calculations [as shown in Fig.~\ref{fig6} (f)]. The IRs of the two bands around the Fermi level near Y point (X point) are $\Gamma_{1}$ and $\Gamma_{2}$, corresponding to opposite $m_y$ ($m_x$) eigenvalue. A Weyl point in 2D can be characterized by the Berry phase $\gamma_{C}$ defined on a small circle $C$ enclosing the point,
 	\begin{equation}
 		\gamma_{C} = \oint_{C} \bm{\mathcal{A}}(\bm{k})\cdot d\bm{k} \quad \mod 2\pi,
 	\end{equation}
 	where $\bm{\mathcal{A}}$ is the Berry connection for the occupied states. One can check that the linear Weyl points in V$_2$STeO under uniaxial compressive strain [as shown in Fig. 6 (e) and (f)] have $\gamma_C=\pi$. 

 It is worth noting that the strain-induced topological semimetal state was not discussed in previous works on V$_2$Se$_2$O and V$_2$SeTeO~\cite{ma2021multifunctional,zhu2023multipiezo}. In this work, we have fully explored the strain-induced topological semimetal state in monolayer materials V$_2$Te$_2$O and V$_2$STeO. Additionally, we found that the strain-induced topological semimetal states in V$_2$Te$_2$O and V$_2$STeO are different: the former shows a nodal loop induced by strain, while the latter exhibits Weyl points. This difference is primarily due to the distinct crystal symmetries of the two materials.
 
%\section{Piezomagnetism}

Strain-induced valley polarization in the monolayer V$_2$Te$_2$O and V$_2$STeO opens up opportunities for generating net magnetization, producing a considerable piezomagnetic effect. When band structure of these materials exhibits valley polarization under applied strain, the Fermi level can be tuned through carrier doping to cross only one valley, producing net magnetic moments. The net magnetization can be defined as $M=\int_{-\infty}^{E_f(n)}\left[\rho^{\uparrow}(\epsilon)-\rho^{\downarrow}(\epsilon)\right] d E$, where $E_f$ is the fermi level after doping, $n$ is the doping density, $\rho^{\uparrow(\downarrow)}$ is the spin-up (down) part of the density of states (DOS), which is dependent on the external strain $\epsilon$. The magnetization under strain and doping concentration (here we focus on hole doping) for monolayer V$_2$Te$_2$O and V$_2$STeO are shown in Fig.~\ref{fig7}(c)-(d) and Fig.~\ref{fig8}(c)-(d), respectively. One observes that the magnetization increases with both carrier density and strain, and the magnetization direction is opposite for tensile and compressive strains. Additionally, in a small strain region, magnetization depends linearly on the strain and eventually saturates for very large strains. The magnitude of the net magnetic moments due to the piezomagnetic effect in monolayer V$_2$Te$_2$O and V$_2$STeO is comparable to those previously reported for antiferromagnetic materials~\cite{ma2021multifunctional,zhu2023multipiezo,jaime2017piezomagnetism,ikhlas2022piezomagnetic}. The doping-dependent band structures of V$_2$Te$_2$O and V$_2$STeO under uniaxial strain are presented in the supplementary material, demonstrating the variation in magnetization polarization. The net magnetic moments induced by the piezomagnetic effect in the V$_2$Te$_2$O and V$_2$STeO monolayers are of the same order of magnitude as those reported for the V$_2$Se$_2$O and V$_2$SeTeO monolayers~\cite{ma2021multifunctional,zhu2023multipiezo}. The distinctive altermagnetic structures of V$_2$Te$_2$O and V$_2$STeO, combined with their low magnetocrystalline anisotropy, allows for effective control of the magnetic direction and moments through methods such as electric field manipulation and doping.

%\section{CONCLUSION}
In conclusion, through ﬁrst-principles calculation and theoretical analysis, we predict four 2D altermagnetic materials--monolayer V$_2$Te$_2$O, V$_2$STeO, V$_2$S$_2$O, and V$_2$SSeO. We demonstrate that these materials are semiconductors with spin-splitting in their nonrelativistic band structures. Their band structures feature a pair of Dirac-type valleys located at the TRIM X and Y points. We demonstrate the effect of strain on the band structure and show that uniaxial strain can induce valley polarization, topological states, and piezomagnetism in these materials. Our results thus reveal this family of 2D altermagnetic materials to be promising platforms for studying the valley physics and for applications in valleytronics, spintronics, and multifunctional nanoelectronics. Experimentally, the 2D altermagnetic structure in real space can be detected by spin-polarized STM/STS, and the momentum-dependent band spin splitting can be directly measured using spin-resolved ARPES. For example, spin splitting in altermagnetic materials MnTe, CrSb and Rb$_{1-\delta}$V$_2$Te$_2$O has been confirmed by ARPES~\cite{krempasky2024altermagnetic, reimers2024direct, zhang2024crystal}. External strain can be induced by depositing these materials onto a substrate with moderate lattice mismatch.

\bigskip
See the supplementary material for band structures of V$_2$Te$_2$O with different hubbard U correction values, band structure with the hybrid functional approach, band structures of V$_2$Te$_2$O and V$_2$STeO with SOC included, the magnetic ground state and stable of the monolayer V$_2$Te$_2$O and V$_2$STeO under the strain, band structures with hole doping for monolayer V$_2$Te$_2$O and V$_2$STeO, and results for V$_2$SSeO and V$_2$S$_2$O. 

\bigskip
%\begin{acknowledgements}
This work is supported by the NSF of China (Grant No. 12204378).
%\end{acknowledgements}

%\bibliography{V2Te2O_ref}
%merlin.mbs apsrev4-1.bst 2010-07-25 4.21a (PWD, AO, DPC) hacked
%Control: key (0)
%Control: author (8) initials jnrlst
%Control: editor formatted (1) identically to author
%Control: production of article title (-1) disabled
%Control: page (0) single
%Control: year (1) truncated
%Control: production of eprint (0) enabled
%

\end{document}